\documentclass[twocolumn,preprintnumbers,amsmath,amssymb]{revtex4-1}

\usepackage[version=3]{mhchem} 
\usepackage{amssymb}
\usepackage{graphicx}
\usepackage{dcolumn}

\begin{document}

\title{EUV ionization of pure He nanodroplets: Mass-correlated photoelectron imaging, Penning ionization and electron energy-loss spectra}

\author{D. Buchta$^1$}
\author{S. R. Krishnan$^2$}
\altaffiliation{Current address: IBM-India Semiconductor R\&D Center, D3-F1, Manyata Park, Bangalore 560045, India}
\author{N. B. Brauer$^3$}
\author{M. Drabbels$^3$}
\author{P. O'Keeffe$^4$}
\author{M. Devetta$^5$}
\author{M. Di Fraia$^6$}
\author{C. Callegari$^7$}
\author{R. Richter$^7$}
\author{M. Coreno$^4$}
\author{K. C. Prince$^7$}
\author{F. Stienkemeier$^1$}
\author{R. Moshammer$^2$}
\author{M. Mudrich$^1$}
\email{mudrich@physik.uni-freiburg.de}

\affiliation{$^1$Physikalisches Institut, Universit{\"a}t Freiburg, 79104 Freiburg, Germany}
\affiliation{$^2$Max-Planck-Institut f{\"u}r Kernphysik, 69117 Heidelberg, Germany}
\affiliation{$^3$Laboratoire de Chimie Physique Mol{\'e}culaire, Swiss Federal Institute of Technology Lausanne (EPFL), 1015 Lausanne, Switzerland}
\affiliation{$^4$CNR Istituto di Metodologie Inorganiche e dei Plasmi, CP10, 00016 Monterotondo Scalo, Italy}
\affiliation{$^5$CIMAINA and Dipartimento di Fisica, Universit{\`a} di Milano, 20133 Milano, Italy}
\affiliation{$^6$Department of Physics, University of Trieste, 34128 Trieste, Italy}
\affiliation{$^7$Elettra-Sincrotrone Trieste, 34149 Basovizza, Trieste, Italy}

\date{\today}


\begin{abstract}
The ionization dynamics of pure He nanodroplets irradiated by EUV radiation is studied using Velocity-Map Imaging PhotoElectron-PhotoIon COincidence (VMI-PEPICO) spectroscopy. We present photoelectron energy spectra and angular distributions measured in coincidence with the most abundant ions He$^+$, He$^+_2$, and He$^+_3$. Surprisingly, below the autoionization threshold of He droplets we find indications for multiple excitation and subsequent ionization of the droplets by a Penning-like process. At high photon energies we evidence inelastic collisions of photoelectrons with the surrounding He atoms in the droplets.
\end{abstract}


\maketitle

\section{\label{sec:Intro}Introduction}
Helium nanodroplets are intriguing many-body quantum systems which feature special properties such as very low equilibrium temperature (0.38\,K), superfluidity, and the ability to efficiently cool and assemble embedded species (`dopants'). Therefore pure He nanodroplets have been extensively studied using electron impact ionization~\cite{Stace:1988,Buchenau:1991,Scheidemann:1993,Halberstadt:1998,Seong:1998} as well as by photoexcitation and ionization with synchrotron radiation~\cite{Froechtenicht:1996,Peterka:2003,Peterka:2007,Haeften:1997,Haeften:2005,Haeften:2011}. Recently, time-resolved experiments have become possible using femtosecond light pulses in the extreme ultraviolet (EUV) spectral range from high-order harmonic generation~\cite{Kornilov:2010,Kornilov:2011,Buenermann:2012,BuenermannJCP:2012}. Based on the photoionization and dispersed fluorescence emission measurements, the following three distinct regimes of excitation and ionization have been identified:

(i) At photon energies $20.5<h\nu<23\,$eV, He nanodroplets are excited with high cross sections into perturbed excited states (``bands'') derived from the 1s2s$^1$S and 1s2p$^1$P He atomic levels. Fast droplet-induced intra-band and inter-band relaxation as well as He$_2^*$ excimer formation follows the excitation~\cite{Haeften:1997,Haeften:2005,Kornilov:2011,Buenermann:2012}. Due to the repulsive interaction between excited He$^*$ or He$_2^*$ and the He environment the excitation migrates to the surface presumably involving both resonant hopping of the electronic excitation as well as nuclear motion of the excited He$^*$ atom~\cite{Scheidemann:1993,Buenermann:2012,BuenermannJCP:2012,Buchta:2013}. Depending on the size of the He droplet, the He$^*$(1s2p$^1$P) state is trapped at the surface and eventually relaxes into the long-lived 1s2s$^{1,3}$S states or into vibrationally excited He$_2^*$ molecules~\cite{Buchenau:1991}. The latter are subject to vibrational relaxation by coupling to the He droplet and eventually evaporate off the droplet surface.

(ii) At photon energies $23<h\nu<24.6\,$eV, the droplet response is even more complex. In addition to the aforementioned relaxation channels, the emission of He$^*$ and He$_2^*$ in Rydberg states dominates~\cite{Haeften:2005,Buenermann:2012,BuenermannJCP:2012}, while the fraction of He$_2^*$ dimers increases with rising excitation energies~\cite{Haeften:1997,Haeften:2005}. At $h\nu>24\,$eV population of triplet states of He was also observed~\cite{Haeften:1997,Kornilov:2010}. As a further relaxation channel, autoionization of He droplets sets in at $h\nu>23\,$eV leading to the formation of small ionic fragments (He$_n^+$, $n\leq 17$) as well as large cluster ions ($N\gtrsim 10^3$)~\cite{Froechtenicht:1996}. A peculiarity of the ionization of He droplets below the ionization energy $E_{i,\mathrm{He}}=24.59\,$eV of atomic He is the emission of electrons with very low kinetic energy $<1\,$meV as seen in photoelectron imaging experiments~\cite{Peterka:2003,Peterka:2007}. Recent time-resolved photoelectron and photoion imaging experiments have revealed the dynamics of various relaxation processes in this regime~\cite{Kornilov:2010,Kornilov:2011,Buenermann:2012,BuenermannJCP:2012}.

(iii) At photon energies $h\nu>24.6\,$eV, that is above $E_{i,\mathrm{He}}$, He$^+$ ions (positive holes) are created in the droplets. The positive charges subsequently migrate through the He droplet by resonant hopping and eventually localize by forming He$_2^+$ molecular ions or by ionizing a dopant if present~\cite{Stace:1988,Scheidemann:1993,Halberstadt:1998,Seong:1998}. The internal energy of the newly formed ion as well as the binding energy liberated upon formation of `snowball' structures (He atoms tightly bound around the ion core) is believed to stop the charge-hopping process and causes massive droplet fragmentation. Therefore, He$^+$ largely from background He atoms and He$_2^+$ from droplets are the dominant species appearing in the mass spectra~\cite{Froechtenicht:1996,Peterka:2007,Kim:2006,Buchta:2013}.

Detailed insight into the dynamics of photoexcitation and ionization of pure He nanodroplets has been gained from ion mass spectra and velocity-map photoelectron imaging~\cite{Peterka:2003,Peterka:2007} as well as from dispersed fluorescence measurements~\cite{Joppien:1993,Haeften:1997,Haeften:2005,Haeften:2011}.
The photoelectron spectra (PES) recorded by ionizing He droplets at $h\nu=25\,$eV have revealed the presence of a high-energy component extending to electron energies $E_e>h\nu-E_{i,\mathrm{He}}$ which was discussed in terms of the direct single ionization of paired up neighboring He atoms to form He$_2^+$ dimer ions in bound vibrational levels~\cite{Peterka:2007}. Photoelectron angular distributions measured for He droplets were found to be more isotropic than those for free He atoms indicating elastic scattering of the escaping electrons from He in the droplets. Apart from electrons created by direct photoionization, electrons with nearly vanishing kinetic energy were observed which arise from an indirect ionization mechanism involving significant electron-He interactions. This component in the PES was most pronounced for large He droplets ionized in regimes (ii) and (iii) up to $h\nu=27\,$eV~\cite{Peterka:2003,Peterka:2007}. Possible origins such as trapping of electrons in so called bubble states that decay when they approach the droplet surface~\cite{Northby:1967,Rosenblit:1995,Haeften:2002,Farnik:2003}, or vibrational autoionization of highly excited electronic states of the droplets were discussed~\cite{Peterka:2007}.

In the present paper we report on a synchrotron study of pure He nanodroplets using Velocity-Map Imaging PhotoElectron-PhotoIon COincidence (VMI-PEPICO) spectroscopy. This method allows us to measure PES and angular distributions in coincidence with specific ion masses which was not possible in previous experiments. The PES and angular distributions  measured in correlation with the most abundant fragments He$_n^+$, $n=1,2,3$ are discussed. We find indications for multiple excitations and subsequent decay by Penning-like ionization when irradiating the droplets at $h\nu=21.6\,$eV which corresponds to the maximum of the 1s$^2\,^1$S$\rightarrow$1s2p$^1$P droplet absorption band~\cite{Froechtenicht:1996,Joppien:1993,Haeften:2001}. Upon ionization of He droplets at high photon energies $h\nu\gtrsim 2\times E_{i,\mathrm{He}}$ we observe low-energy electrons in addition to those directly emitted, which are generated by inelastic electron-He collisions.

\section{Experimental}
The experiments presented here are performed using a mobile He droplet machine attached to a VMI-PEPICO detector at the GasPhase beamline of Elettra-Sincrotrone Trieste, Italy~\cite{OKeeffe:2011}. The experimental setup is described in more detail in a previous publication~\cite{Buchta:2013}. In short, a continuous beam of He nanodroplets with a mean size ranging from 200 to 17000 He atoms per droplet is generated by varying the temperature $T_0$ of a cryogenic nozzle~\cite{Toennies:2004,Stienkemeier:2006}. An adjacent vacuum chamber contains doping cells, which are not used in the experiments reported here unless explicitly mentioned, and a mechanical beam chopper for discriminating ion and electron counts correlated with the droplet beam from background counts due to residual He and other residual gas components.

In the detector chamber further downstream, the He droplet beam intersects the synchrotron light beam at right angle in the center of a velocity map imaging (VMI) spectrometer. The synchrotron radiation is linearly polarized along the direction of the He droplet beam, that is perpendicular to the symmetry axis of the VMI spectrometer. The latter is composed of field plates that accelerate photoelectrons onto a position and time resolving delay-line detector, while photoions are accelerated onto a microchannel plate detector to record flight times. Measuring electrons and ions in coincidence allows us to extract from the data both ion mass spectra and mass-correlated velocity-map photoelectron images. The latter are transformed into PES and angular distributions using standard Abel inversion programs~\cite{Vrakking:2001,Garcia:2004}.

The narrow-band synchrotron radiation ($E/\Delta E\gtrsim 10^4$) is varied between 21 and 66\,eV in this study. All photon energy dependent ion and electron spectra are normalized to the light intensity which is monitored by a calibrated photodiode. Note that a non-negligible amount of second and third order radiation is present at the lower end of the tuning range $h\nu\lesssim 20\,$eV. The pulse repetition rate is 500\,MHz and the peak intensity in the interaction region is estimated to $I\sim 15\,$W\,m$^{-2}$.

An additional beam dump chamber is attached to the end of the apparatus which contains a channel electron multiplier mounted directly in the path of the He droplet beam. It is used for measuring the yield of metastable He atoms and droplets excited by the synchrotron radiation.

\section{Results and discussion}
\begin{figure}
\centering
\includegraphics[width=0.4\textwidth]{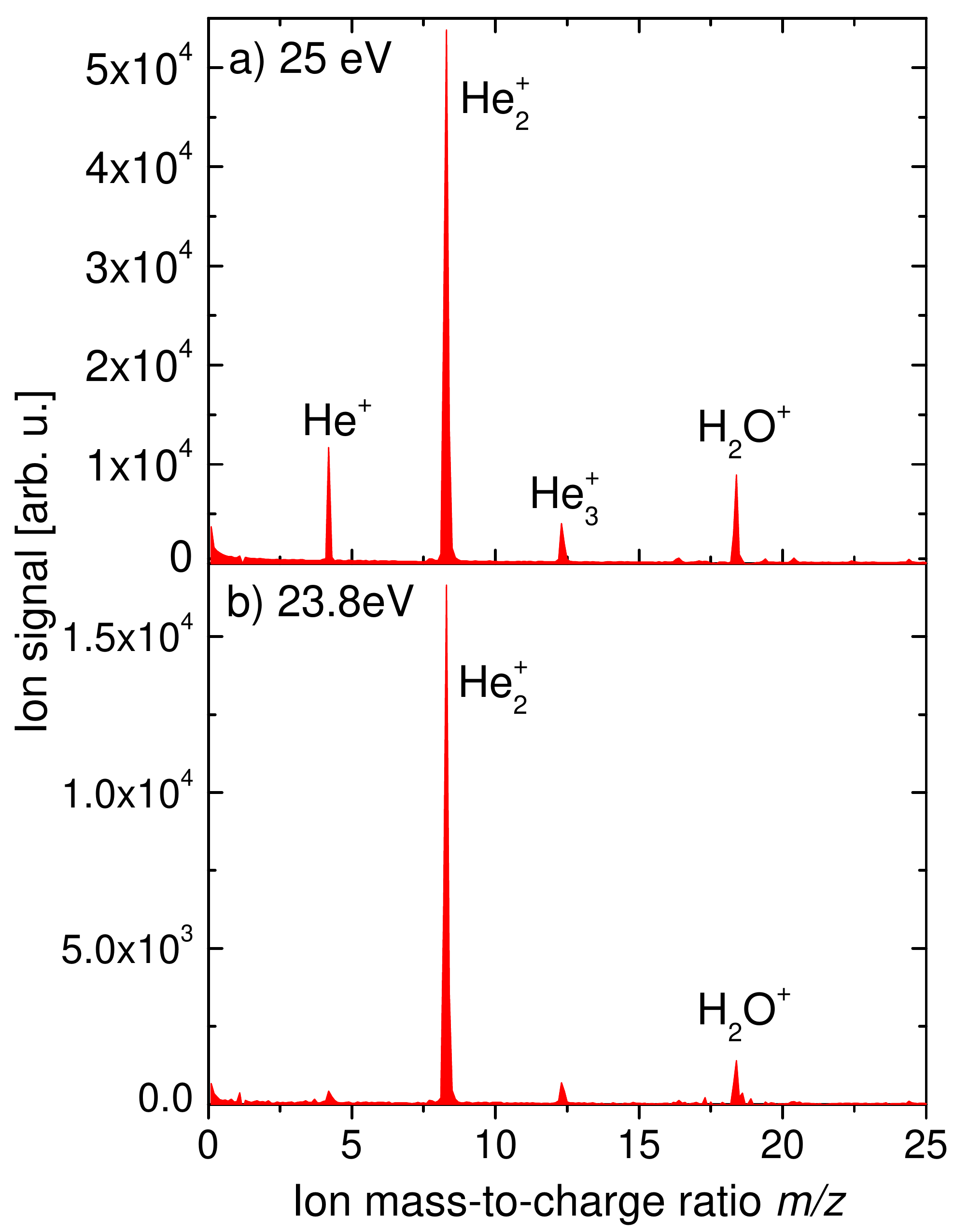}
\caption{Difference mass spectra of ionized (a) and autoionizing (b) He nanodroplets. The He expansion conditions are $p_0=50\,$bar and $T_0=23\,$K ($N=1900$).}
\label{fig:massspectra}
\end{figure}
In this work we focus on ion mass-correlated photoelectron spectroscopy of pure He nanodroplets irradiated by EUV radiation at variable photon energies $h\nu<E_{i,\mathrm{He}}$ up to  $h\nu\gtrsim 2\times E_{i,\mathrm{He}}$. Let us start the discussion of experimental results by presenting typical ion mass spectra. The dependence of electron and ion yields on the experimental parameters (photon energy $h\nu$, He droplet size $N$) will be discussed subsequently. Finally, the ion mass-correlated PES and angular distributions at variable $h\nu$ will be presented.

\subsection{Ion yield spectra}
Fig.~\ref{fig:massspectra} compares typical mass spectra recorded at $h\nu = 25\,$eV (a) and at $h\nu = 23.8\,$eV (b). The use of a mechanical chopper that periodically intercepts the He droplet beam allows us to discriminate the ion signals originating from the He droplet beam from background gas ions. When the beam chopper is in the `open' position, both contributions are measured whereas in the `closed' position, only background ions contribute. Thus, the shown difference signal visualizes the contribution correlated to the He droplet beam only. At $h\nu = 25\,$eV (a) the He atoms in the droplets are directly ionized (regime (iii)), whereas at $h\nu = 23.8\,$eV (b) the He droplets are resonantly excited into the droplet equivalent of the 1s3p$^1$P and 1s4p$^1$P atomic He level out of which they decay by autoionization and other processes (regime (ii)). The He droplet beam source is operated at He expansion conditions of $p_0 = 50\,$bar and $T_0 =23\,$K. The corresponding mean He droplet size amounts to $N = 1900$~\cite{Stienkemeier:2006}.

At photon energies $h\nu > E_{i,\mathrm{He}}$ (Fig.~\ref{fig:massspectra} (a)), the highest mass peaks in the spectra are those of He$^+$ and He$_2^+$. Note that He$_2^+$ is even more abundant than He$^+$, in contrast to earlier electron impact and synchrotron experiments~\cite{Froechtenicht:1996,Buchenau:1991,Peterka:2006,Kim:2006}. This may be due to the long flight distance from the nozzle up to the ionization region of $71\,$cm in our experiment, which results in a highly collimated droplet beam where the content of free He atoms accompanying the droplet beam is suppressed. The efficient formation of He$_2^+$ ions agrees with the established notion that the initially created He$^+$ positive hole migrates within the He droplets and localizes by forming a He$_2^+$ ion. The binding energy liberated by forming the He$_2^+$ molecule as well as by forming a tightly bound shell of He atoms around the ion (`snowball') subsequently induces droplet fragmentation and the ejection of bare He$_2^+$. Higher He$_n^+$ cluster ion masses are also present with lower intensities in the entire mass range shown. The H$_2$O$^+$ signal stems from water molecules picked up by the He droplets from the residual gas which are ionized by charge transfer ionization.

At photon energies $h\nu < E_{i,\mathrm{He}}$ (Fig.~\ref{fig:massspectra} (b)), the He$^+$ signal nearly completely vanishes as expected due to energy conservation. However, at $h\nu =23.8\,$eV, that is in regime (ii), aside from other relaxation channels the excited He droplets are subject to autoionization yielding He$_2^+$ and small He$_n^+$ cluster ions and ultraslow photoelectrons~\cite{Froechtenicht:1996,Haeften:1997,Kornilov:2011,BuenermannJCP:2012}. This explains the high He$_2^+$ yield as compared to all other masses.

\begin{figure}
\centering
\includegraphics[width=0.4\textwidth]{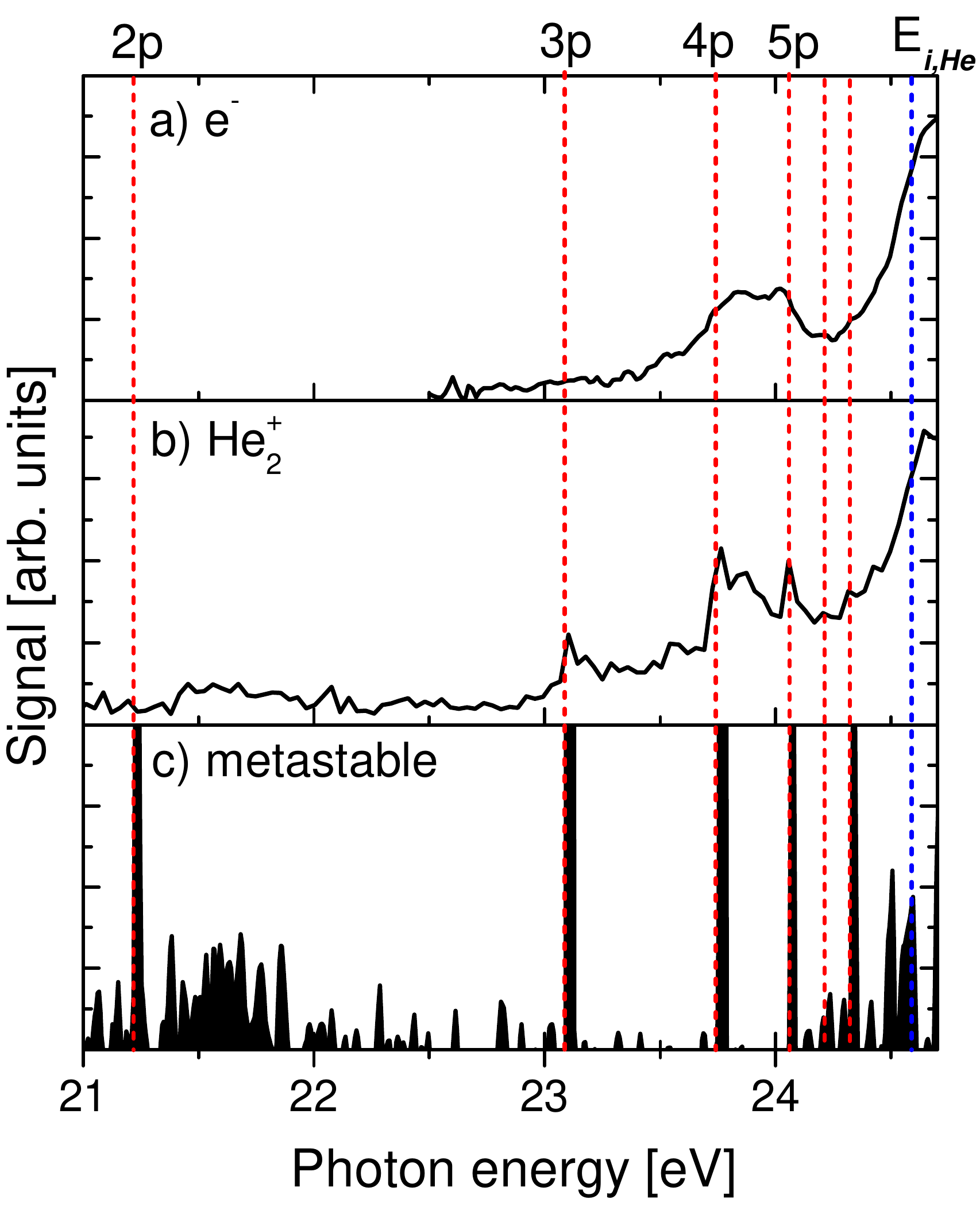}
\caption{Photon energy dependence of the yield of photoelectrons (a), He$_2^+$ ions and metastable atoms and droplets (c). The vertical dashed lines indicate He atomic level energies.}
\label{fig:scans}
\end{figure}
The dependence of the characteristic ionization signals on the photon energy $h\nu$ is studied by recording the electron and He$_2^+$ ion signals for varying $h\nu$. The resulting spectra are depicted in Fig.~\ref{fig:scans}. The He expansion conditions are set to $p_0=50$\,bar and $T_0=21\,$K corresponding to a mean droplet size of $N=2900$.
Different vertical scales are used for the three panels (a)-(c). The yield of metastable atoms and droplets shown in (c) is recorded in the `chopper open' position using a single channel electron multiplier mounted into the He droplet beam. It therefore contains contributions from both the He droplet beam as well as from atomic He effusing into the detection chamber. Note that we cannot strictly exclude contributions to the signal from EUV fluorescence light reaching the electron multiplier.

The broad band structure in Fig.~\ref{fig:scans} (a) and (b) at photon energies $23\leq h\nu\leq 24.6\,$eV is in good agreement with previous ionization spectra recorded with neat He nanodroplets~\cite{Froechtenicht:1996}. Note that we systematically measure higher electron count rates than total ion yields by a factor of 5-15 depending on $h\nu$. This indicates the partial presence of large He$_n^+$ cluster ions with $n>100$ which fall beyond the detection range of our setup~\cite{Froechtenicht:1996}. The peaked structures around $21.8$, $23.1$, $23.8$ and $24.7\,$eV in (a) and (b) can be assigned to excited He droplet states that mostly derive from the 1s2p$^1$P, 1s3p$^1$P, 1s4p$^1$P, and highly excited Rydberg levels of atomic He. At $h\nu > 24.6\,$eV, the He droplets are directly ionized yielding the highest ion and electron signals. When varying the mean droplet size by changing the He nozzle temperature between $T_0=17$ and $27\,$K the overall He$_2^+$ count rate slightly changes with a maximum at $T_0=21\,$K but the structure of the spectrum remains nearly constant. Surprisingly, we find a weak broad maximum in the He$_2^+$ signal around $h\nu = 21.6\,$eV which corresponds to the 1s2p$^1$P excitation band of He droplets. Since this band lies below $E_{i,\mathrm{He}}$ by about 3\,eV, which is more than the binding energy of He$_2^+$, autoionization of singly excited droplets is impossible. As we will discuss below, we attribute this feature to multiply excited He droplets that decay by a Penning-like process in which one He$^*$ excitation relaxes to the ground state whereas the other He$^*$ is ionized.

The signal measured using the ion detector intercepting the droplet beam at the end of the beam line shows sharp peaks corresponding to atomic lines as well as one broad maximum around $h\nu =21.6\,$eV (Fig.~\ref{fig:scans} (c)). The sharp atomic lines at energies corresponding to excitation of the levels 1snp$^1$P, n=2,3,$\dots$ reflect the detection of metastable atomic states of He populated by radiative decay of the $^1$P states excited in the atomic He part of the beam~\cite{Rubensson:2010}.
We attribute the broad peak at $h\nu =21.6\,$eV to the 1s2p$^1$P droplet excitation which decays by relaxation into the metastable 1s2s$^1$S state of He atoms or into the lowest $^1\Sigma_{u,g}^+$ state of He$_2^*$ excited dimers. The latter either remain weakly bound to the droplet surface or desorb off the droplets due to vibrational relaxation~\cite{Buchenau:1991}. Note that the relaxation of the 1s2p$^1$P droplet excitation into 1s2s$^1$S and even lower-lying levels of He$^*$ and He$_2^*$ was previously observed for doped droplets~\cite{Wang:2008,Buchta:2013}.

\subsection{Photoelectron imaging}
\begin{figure}
\centering
\includegraphics[width=0.4\textwidth]{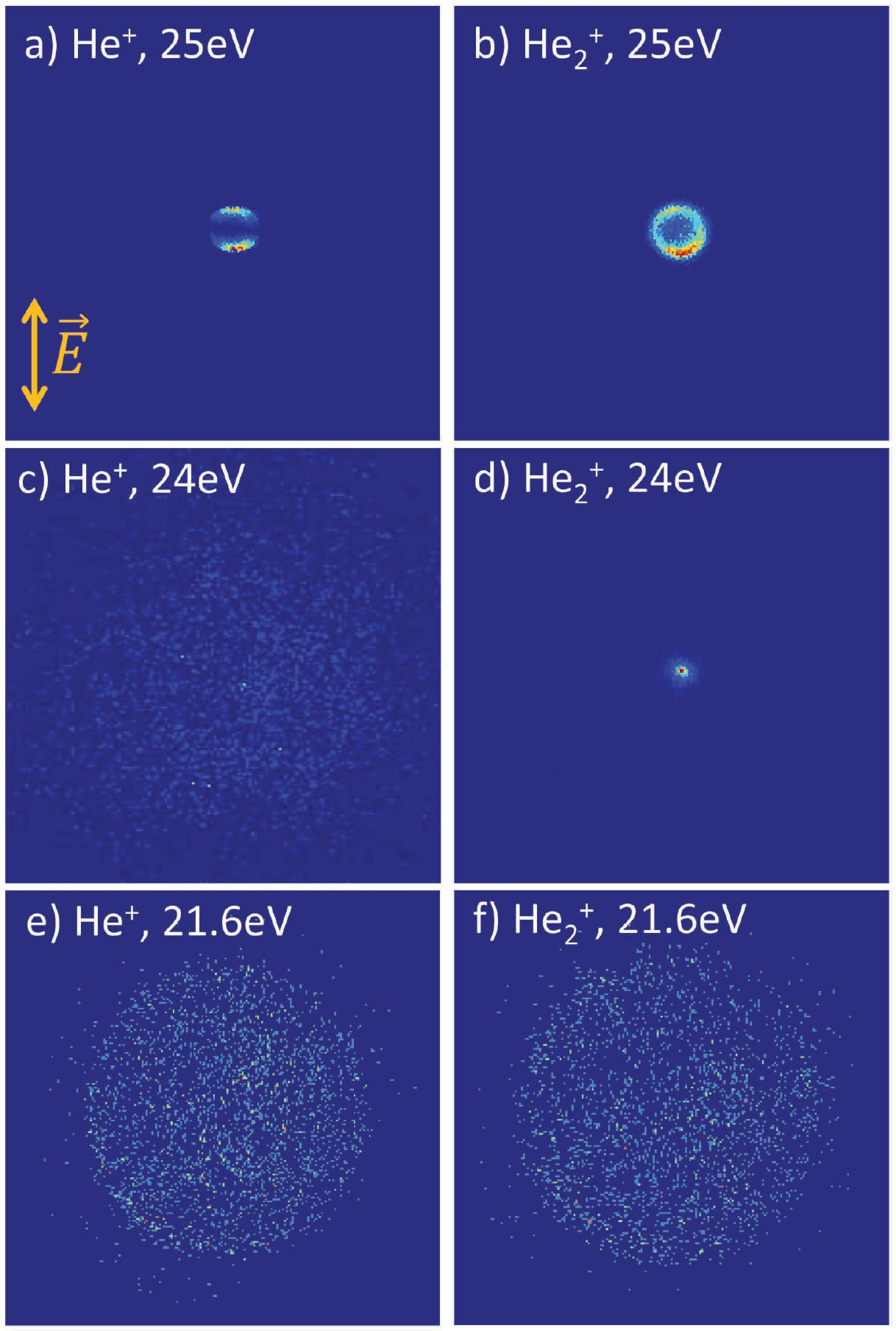}
\caption{Raw velocity map images of photoelectrons from He droplets in correlation with He$^+$ (a), (c), (e) and to He$_2^+$ (b), (d), (f) irradiated at photon energies $h\nu=21.6$, $24$, $25\,$eV. The arrow in (a) indicates the direction of the polarization vector of the EUV radiation.}
\label{fig:VMIHeHe2}
\end{figure}
In order to obtain more detailed information about the dynamics of He droplet ionization in the different regimes (i)-(iii) we record photoelectron images in coincidence with the most abundant ions He$^+$ and He$_2^+$. Fig.~\ref{fig:VMIHeHe2} gives an overview of such images recorded at various photon energies $h\nu$. In these images, the polarization vector of the synchrotron radiation is oriented vertically in the image plane as indicated by the arrow in (a). The electron distribution measured in coincidence with He$^+$ at $h\nu = 25\,$eV (a) shows an anisotropic ring-shaped structure which matches the characteristic angular distribution of a p-wave, as expected for direct one-photon ionization out of the He 1s-orbital. The electrons correlating to He$_2^+$ emitted at the same photon energy (b) feature a similar ring-shaped distribution which has the same radius but is more isotropic. The angular distributions of directly emitted electrons recorded in correlation with He$_2^+$ and He$_3^+$ are analyzed further below.

At the photon energy $h\nu=24\,$eV, that is in regime (ii), the electron signal correlating to He$^+$ nearly vanishes, whereas that correlating to He$_2^+$ concentrates in a small central spot indicating very low electron kinetic energy. As mentioned above, nearly zero kinetic energy electrons have been observed in many experiments with pure and doped He droplets~\cite{Peterka:2003,Peterka:2007,Wang:2008,Kornilov:2011,Fechner:2012,Buchta:2013}. They appear most prominently when $h\nu$ is tuned slightly below $E_{i,\mathrm{He}}$ and droplet autoionization becomes an important decay channel.

Surprisingly, at $h\nu = 21.6\,$eV, that is in regime (i) of pure droplet excitation into the 1s2p$^1$P band, significant electron signals correlating to both He$^+$ and He$_2^+$ are recovered. The two images feature extended isotropic circular structures of nearly equal size. As discussed below in more detail, we attribute these electrons to the decay of multiply excited He droplets by Penning-like ionization.

\begin{figure}
\centering
\includegraphics[width=0.4\textwidth]{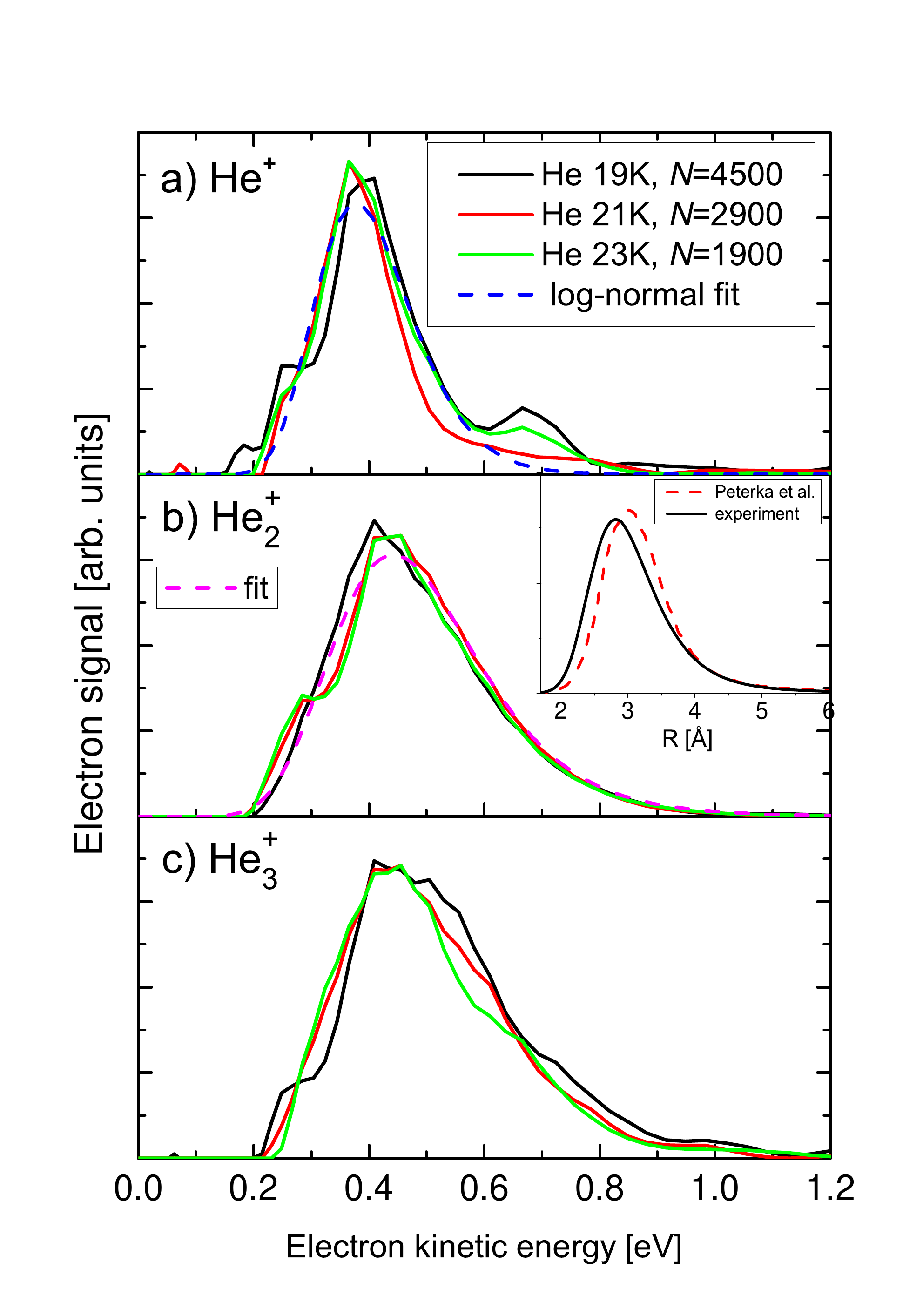}
\caption{Photoelectron spectra measured in coincidence with He$^+$ (a), He$_2^+$ (b), and He$_3^+$ ions (c) at $h\nu=25\,$eV for different He droplet sizes. The He$^+$ spectrum (a) is modeled by a log-normal distribution (dashed line). The dashed line in (b) represents the convolution of the fit function in (a) and a log-normal distribution that accounts for the pure He$_2^+$ spectrum. The inset in b) compares the resulting nearest-neighbor distribution for pairs of He atoms with an ab initio calculation by Peterka et al.~\cite{Peterka:2007}.}
\label{fig:peakform}
\end{figure}
\begin{figure}
\centering
\includegraphics[width=0.3\textwidth]{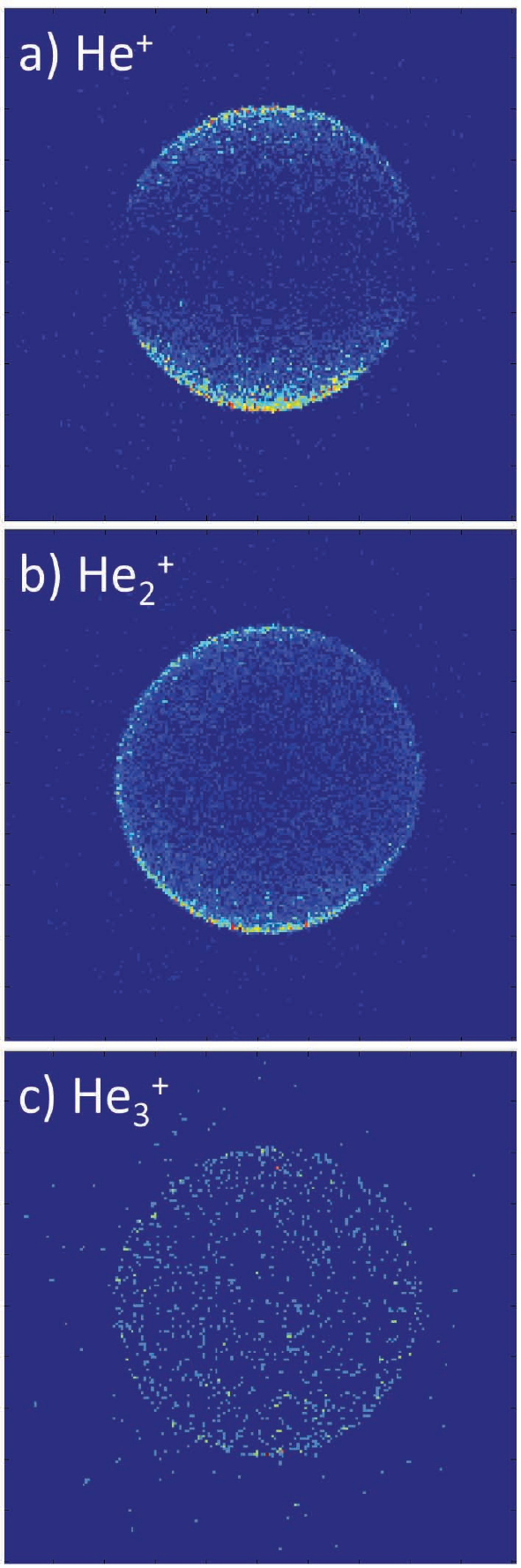}
\caption{Raw velocity map images of photoelectrons from ionized He droplets at $h\nu=35\,$eV recorded in coincidence with He$^+$ (a), He$_2^+$ (b), and He$_3^+$ (c) ions.}
\label{fig:VMI}
\end{figure}
\begin{figure}
\centering
\includegraphics[width=0.4\textwidth]{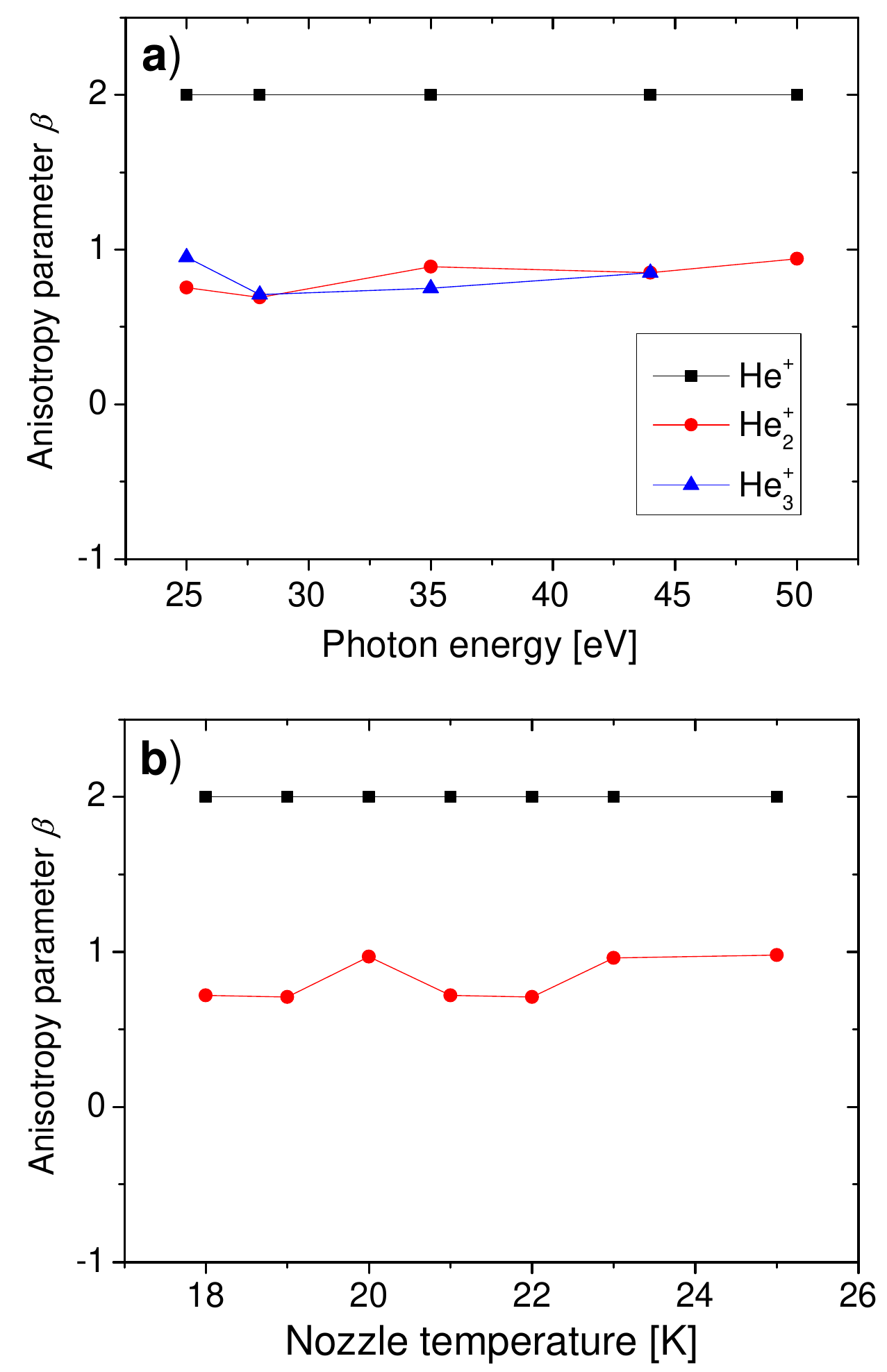}
\caption{Anisotropy parameters $\beta$ inferred from the photoelectron images correlating to He$^+$, He$_2^+$, and He$_3^+$ ions as a function of $h\nu$ (a) and as a function of nozzle temperature in the range $T_0=18$-$25\,$K corresponding to $N=5600$-$1200$ (b). In (a) $N=4500$ and in (b) $h\nu=25\,$eV.}
\label{fig:beta}
\end{figure}

\paragraph{\textbf{Mass-correlated photoelectron spectra}}
First, we examine the PES which we obtain by inverse Abel transformation and angular integration of the photoelectron images recorded in regime (iii). The spectra shown in Fig.~\ref{fig:peakform} are recorded at $h\nu=25\,$eV for different He droplet sizes by varying the nozzle temperature $T_0$ as indicated in the legend. The spectra correlating to He$^+$ (Fig.~\ref{fig:peakform} (a)) are obtained from the background-subtracted (`chopper closed') electron images of the full signal (`chopper open') so as to discriminate the electrons correlated with the He droplet beam. The dashed line represents the result of fitting the average of the experimental curves by a log-normal distribution function, which is and empirical function that well reproduces the measured peak shape. The maximum is peaked at 0.39(2)\,eV which matches the excess energy $h\nu-E_{i,\mathrm{He}}=0.41\,$eV in the direct ionization of He atoms. The width of the spectral feature of He$^+$ reflects the energy resolution of the spectrometer and matches the width measured for atomic He from background gas. Thus, within the experimental uncertainties, we see no significant influence of the presence of He droplets on the PES. This suggests that the He$^+$ atomic ions stem from the atomic component that accompanies the droplet beam. Possibly, there is a contribution from He atoms located at the outer surface of the droplets where the He density is too low to cause significant line shifts and to induce charge migration and He$_2^+$ formation~\cite{Seong:1998}. As discussed in the following section, the high degree of anisotropy of the He$^+$-correlated photoelectrons (Figs.~\ref{fig:VMI},~\ref{fig:beta}) further supports this interpretation.

The peak in the PES correlating to He$_2^+$ (Fig.~\ref{fig:peakform} (b)), however, is slightly shifted and significantly broadened towards higher kinetic energies. The spectrum of He$_3^+$ (c) is even more extended towards higher electron energies than that of He$_2^+$. In the previous work, Peterka et al.~\cite{Peterka:2007} have measured and analyzed the total PES recorded without ion mass correlation. The presence of a shoulder extending to higher energies was interpreted using a simple model based on the assumption that ionization occurs vertically for pairs of closely spaced He atoms in the droplet, thereby accessing the attractive potential region of the cationic He$_2^+$ core. The resulting photoelectrons have higher kinetic energy than those of atomic He~\cite{Peterka:2007}. The shape of this shoulder was simulated by a Franck-Condon model based on the He$_2$ and He$_2^+$ difference potential and the distribution of He-He interatomic distances for the nearest He-He atom pairs (`nearest neighbor') obtained from path integral Monte Carlo calculations.

We adopt the same model in order to infer the distribution $nn(r)$ of distances $r$ for nearest He-He atom pairs which we identify as precursors for He$_2^+$ formation. Note that this model of vertical transitions from He-He pairs into bound levels of He$_2^+$ does not contradict the concept of creating He$^+$ charges that migrate through the droplets before localizing by forming deeply bound He$_2^+$. Due to the presence of one or more additional He atoms close to the respective He-He pair the total interaction potential is extended to a double-well or multiple-well potential where the heights of the barriers depend on the distances between the three or more atoms. For sufficiently close spacing and excitation of high-lying levels above the barriers, the charge is therefore delocalized. Localization then occurs due to vibrational relaxation into one potential well which prevents further hopping over or tunneling through a barrier. Thus, the low-energy edge of the He$_2^+$ peak (Fig.~\ref{fig:peakform} (b))
is associated with He$_2^+$ formed after charge migration, whereas the high-energy tail is identified with direct formation of deeply bound He$_2^+$.

The model curve shown as a dashed line in Fig.~\ref{fig:peakform} (b) is obtained by fitting the convolution of the fit function of the He$^+$ peak (dashed line Fig.~\ref{fig:peakform} (a)) with a log-normal distribution function $f_{He_2^+}(E_e)$. This function is chosen empirically to describe the characteristic line shape of the He$_2^+$ component. From that distribution we obtain $nn(r)$ by mapping the energy distribution $f_{He_2^+}$ onto the shifted difference potential $\Delta V=V_{He_2^+}-V_{He_2}+h\nu-E_{i,He}$~\cite{Janzen:1997,Carrington:1995} using the transformation $nn(r)=f_{He_2^+}(\Delta V(r))d\Delta V(r)/dr$ where $h\nu =25\,$eV. The result is shown as a solid line in the inset of Fig.~\ref{fig:peakform} (b). It only slightly deviates from the original calculation (dashed line) in that its maximum is slightly shifted to a shorter He-He distance $r=2.9\,${\AA} instead of $R=3.0\,${\AA} in the original calculation. Since the measured He$_2^+$ ions are formed in regions of varying density inside or at the surface of the droplets the shown $nn(r)$ distribution is a density average for the present experimental conditions.

\paragraph{\textbf{Mass-correlated photoelectron angular distributions}}
Next, we discuss the angular distributions of photoelectrons correlating to He$^+$, He$_2^+$ and He$_3^+$ in more detail. To this end we record photoelectron images at variable $h\nu$ up to 50\,eV for droplet sizes ranging from $N=1200$ to $5600$. Typical raw photoelectron images recorded at $h\nu=35\,$eV are depicted in Fig.~\ref{fig:VMI}. As for $h\nu=25\,$eV (Fig.~\ref{fig:VMIHeHe2} (a), (b)), we note a reduced anisotropy of the photoelectron distributions correlating to the molecular ions He$_2^+$ and He$_3^+$ (Fig.~\ref{fig:VMI} (b), (c)). From the images, we infer the average anisotropy parameter $\beta$ by fitting the angular dependence of the signal intensity $I(\theta)$ in the inverse Abel transformed images using the standard expression $I(\theta)\propto 1+\beta P_2(\cos\theta)$~\cite{Zare:1972}.

The resulting values of $\beta$ are compiled in Fig.~\ref{fig:beta} (a) for variable $h\nu$ and in Fig.~\ref{fig:beta} (b) for variable $T_0$ (droplet size). While for electrons correlating to He$^+$ we find a constant value $\beta =2.0(1)$ as expected for the direct ionization of unperturbed He atoms, for He$^+_2$ and He$^+_3$ the anisotropy parameter is reduced to $\beta =0.8(1)$ which does not significantly vary as a function of the parameters $h\nu$ and $T_0$. This seems to indicate that He$^+_2$ and He$^+_3$ also merely stem from the He$_2$ and He$_3$ molecular components which accompany the He droplet beam. However, the fact that the photoelectron distributions measured in coincidence with dopant ions generated by charge transfer ionization~\cite{Buchta:2013} strongly resemble those of He$_2^+$ suggests that He$_2^+$ do stem from droplets. The reduced anisotropy is probably due to scattering of the outgoing photoelectron from the He droplets. We rather believe that the probed range of droplet sizes in not sufficiently broad to see a significant influence of a changing average He density on the photoelectron distribution. A better understanding of the photoelectron angular distributions and spectra requires further experimental and theoretical efforts. In particular, the comparison with PES of mass-selected He$_2$ as recently studied~\cite{Havermeier:2010} would give interesting new insight into the effect of the He droplet on the photoelectron distributions.



\paragraph{\textbf{Penning ionization}}
So far we have examined the photoelectron distributions of He nanodroplets in the regime (iii) of direct ionization. However, the photon energy dependent ion yield measurements (Fig.~\ref{fig:scans}) and photoelectron images (Fig.~\ref{fig:VMIHeHe2}) have revealed weak ionization signals even at $h\nu=21.6\,$eV which falls below the droplet autoionization threshold. In these measurements, the He droplets are doped with Na atoms
at a level of about 0.8 atoms per droplet. Under these conditions, Na$^+$ dopant ions are detected as a result of a Penning process where the He$^*$ excitation energy is transferred to the dopant~\cite{Buchta:2013}. However, since we measure the photoelectrons in coincidence with He$^+$ we do not expect any influence of the presence of dopants on the He ionization signals. In particular, the PES measured in coincidence with Na$^+$ ions, discussed in our previous paper~\cite{Buchta:2013}, significantly differ from those correlating to He$^+$ and He$_2^+$. Thus, a false mass-correlation of the latter electrons is excluded.

\begin{figure}
\centering
\includegraphics[width=0.48\textwidth]{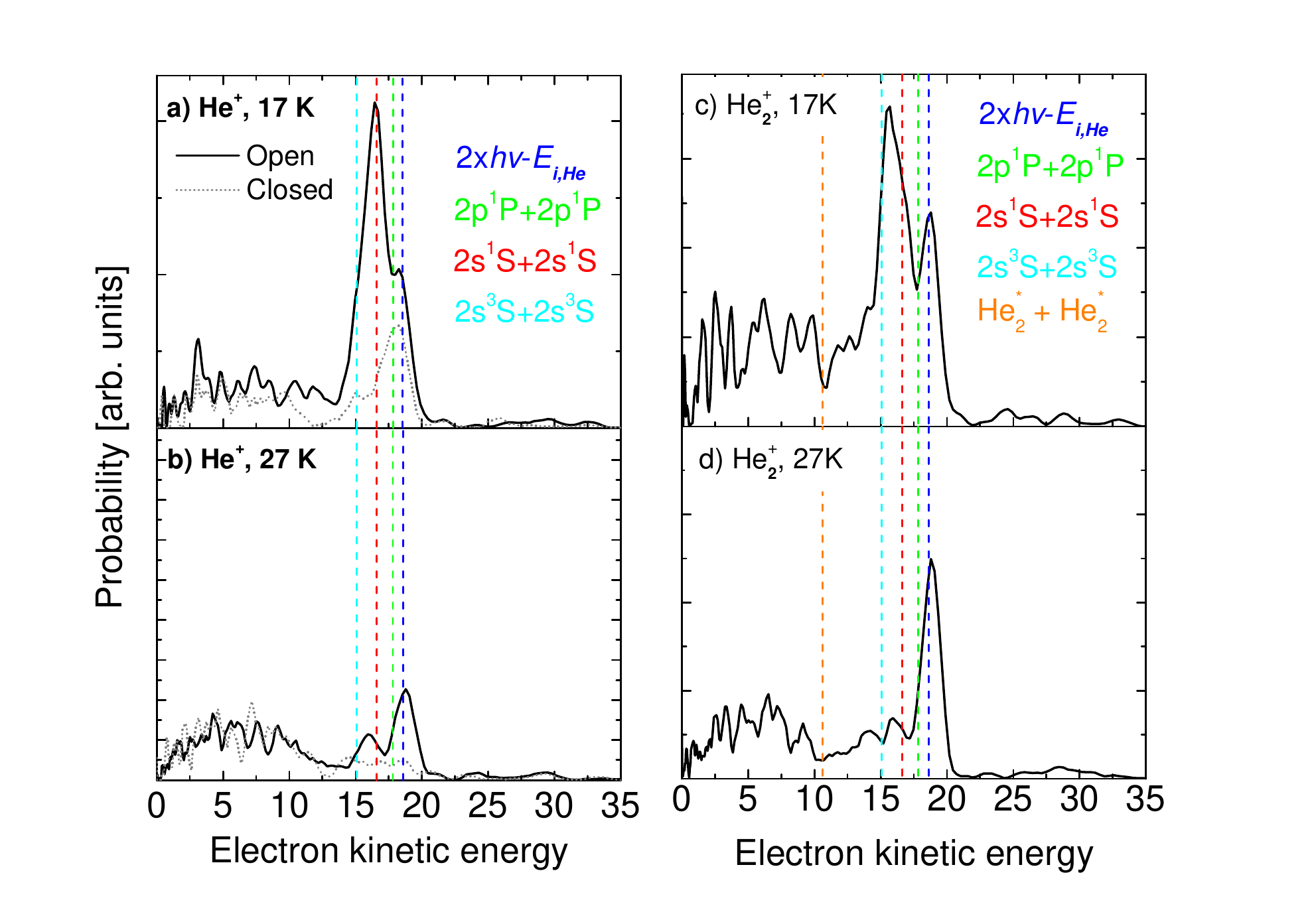}
\caption{Photoelectron spectra measured in coincidence with He$^+$ (a), (b) and to He$_2^+$ ions (c), (d) at $h\nu=21.6\,$eV and at expansion conditions $p_0=50\,$bar and $T_0=17\,$K ($N=7000$) (a), (c) and $T_0=27\,$K ($N=800$) (b), (d). The vertical lines indicate electron energies expected for the relaxation of two excited He atoms into various 1s2p-levels or into the lowest excited state of He$_2^+$. Labels ``Open'' and ``Closed'' refer to data recorded in the open and closed positions of the droplet beam chopper.}
\label{fig:redshift}
\end{figure}

The PES recorded in coincidence with He$^+$ and He$_2^+$ are depicted in Fig.~\ref{fig:redshift} for two different He droplet sizes $N=7000$ ((a) and (c)) and $N=800$ ((b) and (d)). In all measurements we distinguish 3 components in the PES. A sharp peak at electron energies $E_e=2\times h\nu-E_{i,\mathrm{He}}=18.6\,$eV results from direct He ionization by second-order radiation and possibly from electrons emitted by third-order radiation which have undergone an inelastic collision with a surrounding He atoms, as discussed in the last part of this section.
This peak is the dominant feature
in the spectra recorded for small He droplets $N=800$ (Fig.~\ref{fig:redshift} (b) and (d)). In addition, the spectra recorded for large droplets (Fig.~\ref{fig:redshift} (a) and (c)) exhibit a pronounced peak shifted to lower electron energies $E_e\approx 16\,$eV. The broad structure at low energies around $E_e=7\,$eV, which is also present in the spectra from the effusive background, is attributed to background signal presumably from scattered photons and from false coincidences.

These spectra closely resemble the ones measured in coincidence with rare-gas or alkali metal dopant ions~\cite{Wang:2008,Buchta:2013}. In those experiments, the peaked structures similar to the ones visible in Fig.~\ref{fig:redshift} between $h\nu=15$ and $20\,$eV were assigned to electrons produced by ionization of dopants $X$ in a Penning-like process
\begin{equation}
\label{eq:Penning}
\mathrm{He}^*[\mathrm{He}_N] + X \rightarrow \mathrm{He}[\mathrm{He}_N] + X^+ + e^-,
\end{equation}
where the droplet-induced relaxation of He$^*$ into lower-lying levels such as 1s2s$^1$S was likely to precede Penning ionization~\cite{Wang:2008,Buchta:2013}. In the present case of He$^+$ correlated PES, we take the presence of components shifted to lower energies $E_e< 2\times h\nu-E_{i,\mathrm{He}}$ as an indication for a Penning-like reaction involving two He$^*$ excitations in the same droplet as given by Eq.~(\ref{eq:Penning}), for $X=$He$^*$. This process has been discussed in the context of reduced EUV fluorescence emission observed when resonantly exciting large He droplets ($N>10^4$)~\cite{Haeften:2011}.



The probability for double excitation of a He nanodroplet $P_2=P_1^2$ by the interaction with one synchrotron pulse can be estimated from the probability of single excitation, $P_1=N \sigma_a \Delta t I/(e h\nu)\sim 3\times 10^{-9}$, where $I$ is the light intensity. Here, we assume an absorption cross section $\sigma_a\approx 25\,$Mb of one He atom in a droplet containing $N$ atoms, $\Delta t\approx 130\,$ps is the pulse length, and $e$ and $h$ denote the elementary charge and Planck's constant, respectively. When assuming that a fraction of excited He droplets relaxes into metastable states with life times $\tau$ exceeding the pulse repetition period $T=2\,$ns, $P_1$ is replaced by $P_1 t_{tr}/T\approx 500\, P_1$, where $t_{tr}\approx 1\,\mu$s denotes the transit time of the droplets through the focus of the synchrotron beam. For those droplets we obtain $P_2\approx 3\times 10^{-12}$, which yields a signal count rate $S=P_2 N_{HeN}/t_{tr}\sim 0.5\,$s$^{-1}$. Here, the number of He droplets in the focal volume $N_{HeN}$ amounts to $N_{HeN}=n_{HeN} d w^2\approx 6\times 10^4$, where $n_{HeN}=10^8\,$cm$^{-3}$ stands for the number density of He droplets, $d=4\,$mm is the diameter of the droplet beam, and $w=400\,\mu$m is the focus diameter. This estimate approximately matches the count rate measured experimentally. The rapid increase of the Penning signal with increasing droplet size $N$ can be rationalized by the quadratic dependence $S\sim N^2$. Note, however, that this estimate relies on the population of metastable excitations which leads to an accumulation of excitations in one droplet over many light pulses. In contrast to that, He droplets that are multiply excited by single intense ultrashort light pulses as available from free-electron lasers will autoionize by a different mechanism akin to interatomic Coulombic decay~\cite{Kuleff:2010,Mueller:2011}.

Thus, the Penning ionization process appears to be very efficient relative to the decay of He$^*$ or He$_2^*$ excitations by spontaneous relaxation or by desorption off the droplet surface. This interpretation is supported by the results of earlier experiments studying the dynamics of excitations in bulk superfluid He and on molecular beam studies of He$^*$-He$^*$ Penning collisions. In superfluid He the lifetimes of He$^*$ in 1s2s$^{3}$S and of He$_2^*$ excimers in their lowest state $a^3\Sigma_u^+\,(v=0)$ were measured to be about $15\,\mu$s and $13\,$s, respectively~\cite{Keto:1974,Kafanov:2000}. However, upon producing multiple excitations in bulk He the He$_2^*$ excimer population was found to decay due to binary Penning ionization collisions with a rate coefficient rising up to $~2\times 10^{-10}\,$cm$^3$/s at He temperatures $~1.5\,$K~\cite{Keto:1974,Kafanov:2000}. Translating this to a He nanodroplet of radius 50\,{\AA} containing 12000 He atoms and two He$_2^*$ excimers this corresponds to a decay time of about $1.3\,$ns, provided the He$_2^*$ move freely inside the He droplets as in superfluid He. Note that the crossover from diffusive to ballistic motion of He$_2^*$ in superfluid He was observed only recently in the temperature range between 100 and 200\,mK, close to the He droplet temperature~\cite{Zmeev:2013}. The Penning collision rate for He$^*$ can be estimated using the known cross section $\sigma_{\rm{He}^*}\approx 300$~\AA$^2$ from molecular beam scattering experiments~\cite{Mueller:1991}. Assuming a mean relative velocity of the He$^*$ atoms of 60\,m/s the rate coefficient amounts to about $1.8\times 10^{-10}\,$cm$^3$/s and the Penning collision time is $1.5\,$ns. Thus, from these considerations we may expect that Penning ionization occurs within one synchrotron pulse repetition period and inside the interaction region of the droplet beam with the EUV beam. Note that large Penning ionization rates were also observed for He droplets doped with alkali atoms which reside in weakly bound states at the droplet surface~\cite{Scheidemann:1997,Buchta:2013}. Accordingly, Penning ionization of He$^*$ or He$_2^*$ excitations may be enhanced by the simultaneous migration of the two excitations towards the droplet surface due to repulsive He-He$^*$ interaction~\cite{Scheidemann:1993,BuenermannJCP:2012,Buchta:2013} while polarization forces steer them towards each other as in the case of charge migration~\cite{Lewis:2005}.

The vertical dashed lines at $15\leq E_e \leq 18.6\,$eV indicate the energies of electrons created by symmetric Penning reactions of He$^*$ in various electronic levels as denoted in the legend. Other asymmetric combinations of excited states such as He$^*$(1s2p$^1$P)+He$^*($1s2s$^1$S) are also possible but omitted in Fig.~\ref{fig:redshift} for the sake of clarity. Penning ionization of the directly excited 1s2p$^1$P droplet state yields an electron energy $E_e=2\times h\nu -E_{i,\mathrm{He}}$ (blue dashed line) whereas Penning ionization following the relaxation of He$^*$ into atomic levels 1s2p$^3$P, 1s2s$^1$S, 1s2s$^3$S diminishes the Penning electron energy (light dashed lines).

At He expansion conditions where small droplets ($N=800$) are formed (Fig.~\ref{fig:redshift} (b) and (d)), the PES are dominated by direct ionization or by Penning ionization of the unrelaxed He droplet state, which we cannot distinguish.  For larger He droplets $N\gtrsim 4000$ (Fig.~\ref{fig:redshift} (a) and (c)), Penning ionization of He$^*$ that have relaxed into 1s2s$^1$S and lower-lying levels prior to ionization becomes more pronounced. This can be rationalized by the longer migration distances covered by the two He$^*$ excitations in large droplets in order to come close and react. The fact that the PES recorded in coincidence with He$_2^+$ significantly differ from the ones of He$^+$ points at a process where first He$_2^*$ excited dimers form and then the Penning reaction occurs. The conceivable alternative process, He$_2^+$ dimer ion formation after Penning ionization of He$^*$~\cite{Buchenau:1991}, would result in identical PES. The small shift of the Penning peak to lower energies in Fig.~\ref{fig:redshift} (c) indicates that Penning ionization involves He$_2^*$ in various vibronic levels that reach down to even lower energies than the atomic triplet states (vertical dashed line at $E_e\approx 11\,$eV). Given the low statistics of our data and the limited energy resolution of the spectra we cannot infer more details about this ionization mechanism.


\begin{figure}
\centering
\includegraphics[width=0.48\textwidth]{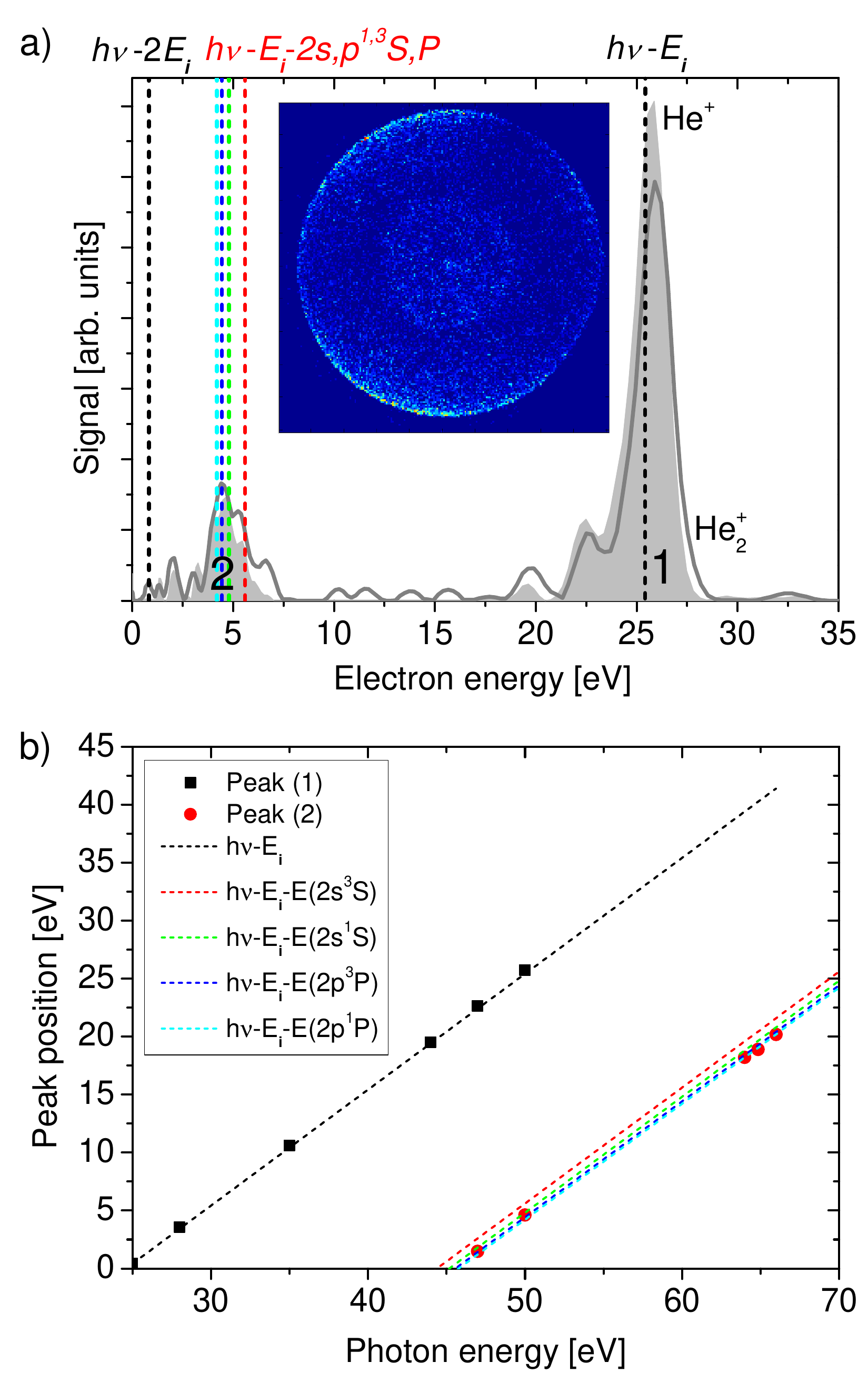}
\caption{(a) Photoelectron spectra recorded in coincidence with He$^+$ (shaded area) and He$_2^+$ (solid line) at $h\nu=50\,$eV ($p_0=50\,$bar, $T_0=23\,$K). Peak 1 corresponds to electrons created directly by ionization of He atoms or He$_2$ dimers in the droplets. Peak 2 stems from electrons that lost energy by inelastic collisions with surrounding He atoms. The inset depicts the photoelectron image correlating to He$_2^+$. (b) Dependence of the peak positions on $h\nu$. The dashed lines depict the energies of photoelectrons emitted directly (black) or after energy-loss by inelastic collisions (colored) when considering the He atomic ionization energy $E_{i,\mathrm{He}}$ and the level energies of the 1s2s$^{1,3}$S and 1s2p$^{1,3}$P atomic levels.}
\label{fig:energyloss}
\end{figure}
\paragraph{\textbf{Inelastic photoelectron-helium collisions}}
Finally, we present PES measured at high photon energies $h\nu >46\,$eV. To the best of our knowledge, no experiments with He nanodroplets at such elevated photon energies have been reported to date. Fig.~\ref{fig:energyloss} (a) depicts typical PES correlating to He$^+$ and He$_2^+$ recorded at $h\nu=50\,$eV. The corresponding raw image for He$_2^+$ is shown as an inset. While the photoelectron distribution correlating to He$_2^+$ is significantly more isotropic than that of He$^+$ (Fig.~\ref{fig:beta}), the PES are nearly  identical at these high photon energies. In addition to the highest peak from directly emitted electrons with energy
$h\nu - E_{i,\mathrm{He}}=25.41\,$eV a second peak appears at energy $E_{E-loss}=h\nu - E_{i,\mathrm{He}} - E_{1s2s,p}\approx 5\,$eV. Here, $E_{1s2s,p}$ stands for the energies of all levels of the 1s2s and 1s2p configurations of He atoms ($^3$S, $^1$S, $^3$P, $^1$P) which can be excited by electron impact but remain unresolved in the PES. This low-energy feature is present at the reduced energy $E_{E-loss}$ in all measured spectra where $h\nu >46\,$eV, as shown in Fig.~\ref{fig:energyloss} (b). It is due to the loss of kinetic energy of the photoelectrons by inelastic collisions with surrounding He atoms as they pass through the droplets in a process of the type
\[
e^-(E_{kin})+{\rm He(1s}^2{\rm )}\rightarrow{\rm He}^* + e^-(E_{kin}-E_{E-loss}).
\]
This interpretation is supported by a vanishing anisotropy parameter $\beta=0.1(3)$ of the angular distribution in the full range of electron energies covered by peak (2) as opposed to $\beta=1.9(1)$ for peak (1). As expected, the information about the direction of emission of the photoelectrons is completely lost by the inelastic electron-He collisions.


The probability $P_{inel}$ for such collisions can be estimated using the well-known inelastic scattering cross sections $\sigma_{inel}$~\cite{Ralchenko:2008}, $P_{inel}=\sigma_{inel}\rho_{He_N}R_{He_N}$. Here, $\rho_{He_N}=0.0218\,${\AA}$^{-3}$ denotes the density of He droplets~\cite{Harms:1998} and $R_{He_N}=(3N/(4\pi\rho_{He_N}))^{1/3}=32\,${\AA} is the droplet radius for a mean droplet size $N=2900$. For $h\nu=50\,$eV we obtain $P_{inel}\approx 6\%$ when summing over all the relevant channels, which roughly matches the ratio of areas of peaks (2) and (1) amounting to about 20\%. When estimating the probability of ionizing He by collisions with photoemitted electrons in the same way, we find $P_{ion}\approx 1$-$2\%$. Unfortunately, this falls below the noise level in our measurements. Note that the elastic electron-He scattering cross section is even larger than the total inelastic cross section for energy-loss collisions by a factor $30$~\cite{Register:1980}. This further confirms our interpretation that the He$^+$ atomic ions, for which we measure a very pronounced anisotropy of the coincident electron distribution, stem from the atomic He component accompanying the He droplet beam.

\section{Conclusion}
Using velocity-map imaging photoelectron-photoion coincidence (VMI-PEPICO) measurements we have investigated the photoionization dynamics of pure He nanodroplets in the regimes of direct ionization and autoionization. We present photoelectron distributions measured in coincidence with the most abundant ion masses He$^+$, He$_2^+$, and He$_3^+$ in a wide range of photon energies. The He$_2^+$ mass-correlated photoelectron spectra are interpreted in terms of contributions from ionized He droplets that relax to form He$_2^+$ and from vertically ionized pairs of nearest neighboring He atoms. The highly anisotropic photoelectron angular distributions recorded in coincidence with He$^+$ indicate that overwhelmingly free He atoms accompanying the droplet beam contribute to the  He$^+$ signal. In contrast, angular distributions of He$_2^+$ and He$_3^+$ display significantly reduced anisotropy, presumably due to scattering of the outgoing photoelectron from the He droplet.

In the regime of pure droplet excitation we measure ionization signals which indicate multiple excitation of the droplets that decay by Penning-like ionization even in the range of small droplets ($N\lesssim 20000$). Future studies at higher photon fluxes as available at free electron lasers will give more insight into the dynamics of multiple and collective excitations. At high photon energies, we observe electron energy-loss processes by inelastic collisions of the photoelectrons with He atoms in the droplets. Such multiple scattering of photoelectrons in clusters is expected to have a severe effect on photoelectron distributions measured in free electron laser experiments.

\begin{acknowledgments}
The authors gratefully acknowledge support by the staff of Elettra-Sincrotrone Trieste for providing high quality light and P. Piseri and C. Grazioli for technical assistance with the VMI-PEPICO detector. Furthermore, we thank
the Deutsche Forschungsgemeinschaft and the Swiss National Science Foundation (Grant no: 200020\_140396) for financial support.
\end{acknowledgments}


%

\end{document}